\documentclass[pra, twocolumn ,amsfonts, amsmath, amssymb]{revtex4}
\usepackage{amsfonts}
\usepackage{amsmath}
\usepackage{bm}
\usepackage{graphicx}
\usepackage{epsf}

\def\pra{{\it Phys. Rev. A}}
\def\prl{{\it Phys. Rev. Lett.}}

\newcommand{\beq}{\begin{equation}}
\newcommand{\eeq}{\end{equation}}
\newcommand{\beqa}{\begin{eqnarray}}
\newcommand{\eeqa}{\end{eqnarray}}

\newcommand{\ket}[1]{| #1    \rangle }
\newcommand{\bra}[1]{ \langle   #1  | }

\newcommand{\amp }[2]{ \langle #1 |  #2  \rangle }

\begin{document}

\title{Not So SuperDense Coding - \\ Deterministic Dense Coding with Partially Entangled States}
\author{Shay Mozes$^{(1)}$,
Benni Reznik$^{(1)}$, Jonathan Oppenheim$^{(2)(3)}$ }
\affiliation{$^{{1}}$ Department of Physics and Astronomy,
Tel-Aviv University, Tel Aviv 69978, Israel. }

\affiliation{$^{(2)}$
Dept. of Applied Mathematics and Theoretical Physics, University of
Cambridge, Cambridge, U.K.}

\affiliation{$^{(3)}$ Racah Institute of Theoretical Physics,
Hebrew University of Jerusalem, Givat Ram, Jerusalem 91904, Israel}

\date{\today}
\begin{abstract}
The utilization of a $d$-level partially entangled state, shared by two
parties wishing to communicate classical information without
errors over a noiseless quantum channel, is discussed. We
analytically construct deterministic dense coding schemes for certain
 classes of non-maximally entangled states, and numerically obtain schemes
in the general case. We study the dependency of the information
capacity of such schemes on the partially entangled state shared
by the two parties. Surprisingly, for $d>2$ it is possible to have
deterministic dense coding with less than one ebit.
In this case the number of alphabet letters that can be
communicated by a single particle, is between $d$ and $2d$.
In general we show that the
alphabet size grows in ``steps"  with
the possible values $ d, d+1, \dots, d^2-2 $. We also find that
states with less entanglement can have greater communication
capacity than other more entangled states.
\end{abstract}

\maketitle

\section{Introduction}
Dense coding, originally introduced by Bennett and Wiesner
\cite{BW92} is the surprising utilization of entanglement to
enhance the capacity of a quantum communication channel. Two parties,
Alice and Bob, communicate by sending
a qubit over a noiseless quantum channel. As
no more than two spin states can be perfectly distinguished, Alice
can encode only one of two different letters, say '0' or '1',
within each particle she sends. This is no better than using a
classical communication channel. However, Bennett and Wiesner have
shown that if Alice and Bob each hold one particle of a maximally entangled
pair, it is possible for the sender, Alice, to transform the
two-particle state into 4 orthogonal states by acting locally on
her particle. After sending Bob her half of the pair, he will be
able to perfectly distinguish the one of four different states by measuring
the pair of particles collectively. Surprisingly, this enables the
transmission of one of four letters by sending a single
qubit, provided that the two parties share
initial entanglement.

Numerous aspects of dense coding have been studied. Among these are
generalization to pairs of entangled $d$-level systems \cite{BW92},
to continuous variables
\cite{BK00} and to settings involving more than two parties \cite{BVK98}.
Other works \cite{BE95,capacity,BVP00}
studied dense coding in the asymptotic limit, where many copies
of a partial entangled state are used.

In this paper we consider the case of pure non-maximal entanglement
shared between a pair of separated $d$-level systems.
We are not interested in the asymptotic channel capacity, but rather in the
\emph{deterministic} procedure, where the parties wish to perfectly distinguish
messages encoded on a single $d$-level particle. We use exact and numeric
methods to study the relation between the given state $\psi$ whose
entanglement is given by its entropy, $S(\psi)$, to $N_{max}(\psi)$, the maximal size
of alphabet which can be perfectly transmitted. In other words,
 $N_{max}(\psi)$ denotes the maximal number of orthogonal states which
 can be generated by means of a unitary transformation acting  locally
 on one side of the given state.

Our results suggest that for $d>2$, deterministic dense coding processes
which utilize non-maximally entangled states
are possible for an alphabet size, $N_{max}(\psi)$, changing in
``steps": $N_{max}(\psi) \in\lbrace d, d+1, \dots, d^2-2\rbrace $
(see Fig. 1).
Interestingly, the last step with $d^2-1$ letters seems to be missing.
We have been able to demonstrate these results analytically
when $N_{max}$ is a multiple of $d$, $N_{max}(\psi)=kd$, $k=2\dots d-1$,
and for the special case $N_{max}(\psi) =d+1$.
Using numerical methods, the existence of the other steps
has been fully verified for $d=3$ and $d=4$, and partially in higher dimensions.

We have further computed the minimal entanglement $S_{min}(\psi)$ required to
obtain $N$ letters. For  $N\le 2d-1$, the minimal required entanglement
turns out to be less than one ebit (see Fig. 2).
Therefore, our method is not equivalent to the trivial approach
wherein deterministic concentration brings a non-maximal state
to an ebit \cite{popescu}, to be  used
by utilizing the standard dense coding scheme.

In addition, we find that entanglement, while playing an important role
in the communication capacity, does not completely determine
$N_{max}(\psi)$.
We show that one can have two states with the
property that the less entangled one is in fact better for communication.
That is, we can have $N_{max}(\psi_1)>N_{max}(\psi_2)$ while
$S(\psi_1)<S(\psi_2)$.
This is perhaps reminiscent of \cite{HSSH02} where it was shown that states
with less entanglement can sometimes have a greater probability of being
distinguished by separated parties who can only communicate classically.
A related situation has been reported in \cite{reznikpovm}
wherein non-maximal states, rather then maximal,
were needed to perform certain remote operations.

This paper is organized as follows.
We first review deterministic dense coding with
maximally entangled $d$-level systems. Then we proceed to
formulate the problem considered in this paper. Section \ref{2D} treats
the two-dimensional case analytically and shows that non-maximal states
cannot be used to distinguish more than two letters.
In section \ref{geometric} we present a
geometric approach, to construct steps with $N_{max}=kd$.
There we show that one can have more communication with less entanglement.
In section \ref{general}, a more
general though less intuitive analytic approach is presented, followed
by our numerical results.
Finally, we summarize our results in section VII.

\section{Dense Coding with Maximal Entanglement}
We consider a bipartite qu$d$it pure state. That is, a system composed of two $d$-level separated
subsystems. This system is initially prepared in a maximally entangled state:
\beq
\ket{\psi_{00}}=\frac{1}{\sqrt{d}} \sum_{i=0}^{d-1} \ket{i}_A \otimes \ket{i}_B
\eeq
where A (B) denotes Alice's (Bob's) subsystem.
Alice encodes an alphabet of size $d^2$ which we denote as $\{(m,n)\}_{m,n=0}^{d-1}$ using a set
$\{U_{mn}^A\}$ of local unitary operations on particle A.
There are many possible \cite{werner} realizations for this set of operators.
An elegant, and undoubtedly the most common construction is
\beq
U_{mn} = (X)^m(Z)^n
\eeq
where $X$, the \emph{shift} operator and $Z$, the \emph{rotate} operator are defined by:
\beqa \label{eq:shiftnrotate}
X \ket{k} & = & \ket{(k+1)_{mod(d)}} \nonumber \\
Z \ket{k} & = & e^{\frac{2 \pi i k}{d}} \ket{k}
\eeqa
It can be easily verified that $\ket{\psi_{mn}} = (U_{mn} \otimes \openone_B) \ket{\psi_{00}}$ form an orthogonal basis of the two
qu$d$its Hilbert space. After encoding the letter $(m,n)$, Alice sends her particle to Bob through the
quantum channel. Bob performs a projective measurement of the two-particle state on $\{\ket{\psi_{mn}}\}$ to decode the message.

A few remarks are in order here.
First, we note that for qubits ($d$=2), this basis is just
the well known Bell basis:
\beqa
\ket{\psi_{00}} = \frac{1}{\sqrt{2}}(\ket{00}+\ket{11}) \nonumber \\
\ket{\psi_{01}} = \frac{1}{\sqrt{2}}(\ket{00}-\ket{11}) \nonumber \\
\ket{\psi_{10}} = \frac{1}{\sqrt{2}}(\ket{10}+\ket{01}) \nonumber \\
\ket{\psi_{11}} = \frac{1}{\sqrt{2}}(\ket{10}-\ket{01})
\eeqa
Second, trying to intuitively understand the difference between the classical
and quantum cases, we note
that the \emph{shift} operators may be regarded as classical, in the sense that they correspond to the
possibility of sending $d$ distinct values of a classical $d$it.
The \emph{rotate} operators may be regarded as the quantum enhancement, which enables the local realization
of $d^2$ orthogonal $d=2$ qu$d$its states.

\section{Deterministic Dense Coding with Non-Maximal Entanglement}
We now introduce the main problem this paper addresses. Instead of using a maximally entangled
state, we consider an arbitrary pure state.
This can be written in the Schmidt \cite{Peres} representation as:
\beq \label{eq:psi}
\ket{\psi}=\sum_{i=0}^{d-1} \sqrt{\lambda_i} \ket{i}_A \otimes \ket{i}_B
\qquad ; \qquad\sum_{i=0}^{d-1} \lambda_i = 1
\eeq
where $\ket{i}_A$ ($\ket{i}_B$) are the Schmidt basis for system A (B).

We are interested in a maximally sized set of local unitary operators
$ \{ U_i^A\}_{i=1}^{N_{max}(\psi)} $
that maintain orthogonality. That is, for all $1 \le i,j \le N_{max}(\psi)$ we have:
\beq \label{eq:orth1}
 \bra{\psi}(U_i^\dagger \otimes \openone)(U_j \otimes \openone) \ket{\psi} = \delta_{i,j}
\end{equation}
Substituting the state (\ref{eq:psi}) into (\ref{eq:orth1}) yields:
\beqa \label{eq:orth2}
\delta_{i,j} &=& \sum_{k,l=0}^{d-1} \sqrt{\lambda_k \lambda_l} _A \bra{k} U_i^\dagger
U_j \ket{l}_A \otimes _B\amp{k}{l}_B \nonumber \\
&=& \sum_{k=0}^{d-1} \lambda_k \bra{k} U_i^\dagger U_j \ket{k} \nonumber \\
&=& Tr( \Lambda U_i^\dagger U_j )
\eeqa
where $\Lambda$ is a $d \times d$ diagonal matrix of the Schmidt coefficients
($\Lambda_{ii}=\lambda_i$).
Note that the matrices $U_i$ are unitary in the usual sense
$U_{i}^\dagger U_i=\openone$,
but the orthogonality is defined with respect to a non-trivial weights vector
(the Schmidt coefficients) rather than
the usual trace. For the rest of this paper orthogonality should be understood in this sense.

In this paper our goal is to study the effect of the initial state $\ket{\psi}$ on the
maximal size $N_{max}(\psi)$ of the set of unitaries satisfying (\ref{eq:orth2}).
Put in other words, we would like to understand and characterize the relation $N_{max}(\psi)$.

We first note that for any choice of $\psi$,
there always exists a set of at least $d$ such unitaries.
This is the set of \emph{shift} operators introduced in the previous section.
Let us explicitly verify that orthogonality is indeed maintained:
\beqa
&{}& \bra{\psi}({X^n}^\dagger \otimes \openone) (X^m \otimes \openone) \ket{\psi}  =\nonumber \\
&=& \sum_{i,j}\sqrt{\lambda_{i}\lambda_{j}}\amp{(i+n)_{mod(d)}}{(j+m)_{mod(d)}} \otimes \amp{i}{j} \nonumber \\
&=&  \delta_{n,m}
\eeqa
That this set is always orthogonal should not surprise us as it corresponds to
the classic possibility of encoding $d$ distinct values in a single $d$it.

\section{The Two-Dimensional Case} \label{2D}

We first consider the case of non-maximally entangled qubits ($d=2$).
We shall show that for all non-maximally entangled states,
only $N_{max}(\psi)=2$ unitaries can be constructed.
This means that \emph{deterministic} dense coding with partial entanglement
is not possible in $d<3$ dimensions; partially entangled qubits have no
advantage over pure product states or classical bits.

For convenience, and without loss of generality we assume that $\openone \in \{U_i\}$.
We parameterize $U = e^{i \vec{\sigma} \cdot \hat{n} \theta} = \cos{\theta} \openone + i \sin{\theta} (\vec{\sigma} \cdot \hat{n})$
, where $\vec{\sigma}$ are the Pauli matrices, and $\hat{n}$ is a unit vector.
Since $\openone \in \{U_i\}$ we must have for all $U \in \{U_i\}$, $Tr(\Lambda U)=0$.
That is:
\beq
0 = (\lambda_0 + \lambda_1)\cos{\theta} + i(\lambda_0  - \lambda_1) \hat{n}_z \sin{\theta}
\eeq
which determines $\theta = \frac{\pi}{2}$ and $\hat{n}_z=0$. Suppose we want to have a set of three
unitaries $\{\openone , U_1 , U_2 \}$. $U_{1(2)}$ must therefore be of the form
$U_{1(2)}= i(\sigma_x x_{1(2)} + \sigma_y y_{1(2)})$. We must also satisfy:
\beqa \label{eq:dim2}
0 & = & Tr(\Lambda U_1^\dagger U_2) \nonumber \\
& = & (\lambda_0 + \lambda_1)(x_1x_2+y_1y_2) + i(\lambda_0 - \lambda_1)(x_1y_2-y_1x_2) \nonumber \\
\eeqa
For non-maximal entanglement we have $\lambda_0-\lambda_1 \neq 0$ and from the normalization we
also have $\lambda_0 + \lambda_1 = 1$. In addition we have $x_1^2 + y_1^2 = x_2^2 + y_2^2 = 1$.
Combining all these restrictions (\ref{eq:dim2}) has no solutions.

\section{Higher dimensions, the geometric approach} \label{geometric}
Regarding the \emph{shift} operators as ``classical'', and the \emph{rotate} operators as the
quantum enhancement made possible by the entanglement of the initial state, one may try to
generalize the dense coding scheme by constructing rotations, or \emph{phase} operators suitable for
the given non-maximal entanglement. In analogy to (\ref{eq:shiftnrotate}), we are looking for a
set $\{ {Z_i} \}_{i=1}^N$
defined by:
\beq
Z_n\ket{k} = e^{i\theta_k^n}\ket{k}
\eeq
where $\theta_k^n$ are real phases whose choice will be discussed shortly.
The orthogonality requirement dictates:
\beqa \label{eq:geom1}
0 &=&\bra{\psi}Z_n^\dagger Z_m \ket{\psi} \nonumber \\
& = & \sum_{i,j}\sqrt{\lambda_{i}\lambda_{j}}e^{i(\theta_j^m-\theta_i^n)}\amp{ii}{jj} \nonumber \\
& = & \sum_{i} \lambda_ie^{i(\theta_i^m-\theta_i^n)}
\eeqa
We can use a set of $N$ such operators to construct $Nd$ orthogonal operators
(in the sense of (\ref{eq:orth2})), namely $U_{mn} = (X)^m(Z_n)$ where $0 \leq m < d$ and
$0 \leq n < N \leq d$.
This construction implies that the total number of operators is a multiple of $d$.
In the classical or non-entangled case, it is $1 \cdot d$, and in the maximal case it is $d \cdot d$.
To see the effect of the initial state $\ket{\psi}$ on $N_{max}(\psi)$, let us examine the simple case
where we look for $N=2$ phase operators.
Again, we assume that $\openone \in \{Z_i\}$, so (\ref{eq:geom1})
reduces to $\sum_i \lambda_i e^{i\theta_i}=0$. In other words, we are faced with the geometric
task of forming a polygon using $d$ vectors of lengths $\{\lambda_0,\lambda_1,\dots,\lambda_{d-1}\}$.
This can always be accomplished if the longest vector is shorter than the sum of the others.
Assuming that the $\lambda$'s are sorted in descending order this condition is simply
$\lambda_0 \leq 1/2$. Not surprisingly, this corresponds to entanglement $S(\psi) \geq 1$.
Note that all states with $\lambda_0 \leq 1/2$ are majorized by a Bell state
in a $d$-dimensional Hilbert space ($ \frac{1}{\sqrt{2}}\ket{00} +
\frac{1}{\sqrt{2}}\ket{11} + 0\sum_{i=2}^{d-1}\ket{ii}$), and thus can be converted to it by LOCC \cite{popescu,nielsen}.
For the Bell state a
construction similar to (\ref{eq:shiftnrotate}) trivially yields $2d$ orthogonal states.
Note, however, that in order to deterministically concentrate
 $\ket{\psi}$ into a Bell state one must use
both local operations (LO) and classic communications (CC) \cite{popescu,MK01,nielsen},
whereas in our construction only LO are used.  Use of additional communication to
convert non-maximally entangled states to maximally entangled ones, would reduce
the net gain in communication.

For the case where we want $N>2$ \emph{phase} operators satisfying (\ref{eq:geom1}),
we have not found a
simple geometric interpretation.
Similar phase factors are also used in the context of deterministic teleportation schemes \cite{G04}.
It can be shown that such phases can be found if, and only if, $\lambda_0 \le 1/N$.
For example, for $d=4$ and $N=3$, we can construct $4 \cdot 3 = 12$ operators when
$\frac{1}{3}=\lambda_0 = \lambda_1 = \lambda_2+\lambda_3$ by regarding
$\frac{1}{\sqrt{3}} \left( \ket{00} + \ket{11} +  \sqrt{3}(\sqrt{\lambda_2}\ket{22} + \sqrt{\lambda_3}\ket{33}) \right)$
as a maximally entangled state in three dimensions, so we can use the powers of the operator $Z$ in (\ref{eq:shiftnrotate})
with $d=3$ as $3$ phase operators.

These constructions, as well as the numerical results (Fig. 1) of
the following section, lead to the conclusion that in finite dimensional systems, $N_{max}$,
The maximal number of
orthogonal unitaries, does not depend directly on the
entanglement, but on some other function of the
coefficients ${\lambda_i}$. We can have
$S(\psi_1) > S(\psi_2)$, but $N_{max}(\psi_1) < N_{max}(\psi_2)$.
Naively, one may have expected
more entanglement to mean greater communication capacity, yet this is not so.

\section{A General Approach} \label{general}
It turns out that the geometric approach, although guided by the appealing
separation into ``classical''
and quantum operators, does not provide the most general construction.
Consider the initial state
\beq
\ket{\psi_3} = \sqrt{\frac{2}{3}}\ket{00} + \sqrt{\frac{1}{3}}\ket{11}+ 0\ket{22}
\eeq
in $d=3$ dimensions. Since $\lambda_0 = \frac{2}{3} > \frac{1}{2}$,
using the results of the last section, one might be
tempted to conclude that the maximal size of a set of orthogonal unitaries is just $d=3$. But if we abandon the
 \emph{phase} and \emph{shift} operators, one may consider the
set $\{ \openone, X, U_3, U_3^\dagger \}$
,where
\beq
U_3 =  \left( \begin{array}{ccc}
-\frac{1}{2} & 0 & -\frac{\sqrt{3}}{2} \\
0 & 1 & 0 \\
\frac{\sqrt{3}}{2} & 0 & -\frac{1}{2} \end{array} \right)
\eeq
is a rotation by $\frac{2\pi}{3}$ within the subspace spanned by
 $\{\ket{0}, \ket{2} \}$. This set consists of four orthogonal unitaries
(with respect to $\ket{\psi_3}$), and as will be discussed in
the sequel, $\ket{\psi_3}$ is the state with minimal entanglement in $d=3$
dimensions admitting more than three
orthogonal unitaries.
Note that the above construction is by no means unique.
It can be generalized to arbitrary dimension $d$ as follows:
The partially entangled state is $\mathbb{C}^d \otimes \mathbb{C}^d \ni \ket{\psi_d} = \sqrt{\frac{d-1}{d}}\ket{00} + \sqrt{\frac{1}{d}}\ket{11}$, and the set of $d+1$
orthogonal unitaries is $\{ \openone_d, X \} \cap \{U_d^k\}_{k=0}^{d-2 }$, where
\beqa
U_d^k\ket{0} & = & -\frac{1}{d-1}\ket{0} + \frac{\sqrt{d}}{d-1}\sum_{j=1}^{d-2}
e^{\frac{2\pi i k j}{d-1}}\ket{j+1} \nonumber \\
U_d^k\ket{1} & = & \ket{1}
\eeqa
The effect of $U_d^k$ on all other basis vectors is restricted only by the
unitarity requirement ${U_d^k}^\dagger U_d^k  =  \openone$.
Let us verify explicitly that $\{U_d^k\}$ is indeed an orthogonal set (we ommit the subscript $d$):
\beqa
trace(\Lambda {U^k}^\dagger U^l) & = & \frac{d-1}{d}\bra{0}{U^k}^\dagger U^l)\ket{0}
+ \frac{1}{d}\bra{1}{U^k}^\dagger U^l)\ket{1} \nonumber \\
& = &\frac{1}{d(d-1)} \left( 1 + d\sum_{j=1}^{d-2}
e^{\frac{2\pi i (l-k) j}{d-1}} \right) + \frac{1}{d} \nonumber \\
& = & \frac{1}{d-1} + \frac{1}{d-1}\sum_{j=1}^{d-2}e^{\frac{2\pi i (l-k) j}{d-1}}
\nonumber \\
& = &\frac{1}{d-1}\sum_{j=0}^{d-2}e^{\frac{2\pi i (l-k) j}{d-1}}
 = \delta_{k,l}
\eeqa
and trivially
\beqa
trace(\Lambda \cdot \openone \cdot U^k) & = &\frac{d-1}{d}\bra{0}{U^k}\ket{0}
+ \frac{1}{d}\bra{1}{U^k}\ket{1} \nonumber \\
& = & -\frac{d-1}{d} \frac{1}{d-1} + \frac{1}{d} = 0 \nonumber \\
trace(\Lambda X^\dagger U^k) & = &\frac{d-1}{d}\bra{0}{X^\dagger U^k}\ket{0}
+ \frac{1}{d}\bra{1}{X^\dagger U^k}\ket{1} \nonumber \\
& = & \frac{d-1}{d}\bra{1}{U^k}\ket{0}
+ \frac{1}{d}\bra{2}{U^k}\ket{1} = 0\nonumber \\
\eeqa
This construction can be further generalized to any case where
$\lambda_1 = \frac{d}{N} = \frac{m-1}{m}$ for some integer $m$.
Note that for large $d$, $\lambda_1 = \frac{d-1}{d} \simeq 1$, which means that the entanglement required for
having more than the ``classical'' $d$ unitaries approaches zero.

\begin{figure}
\begin{center}
\includegraphics[width=0.45\textwidth]{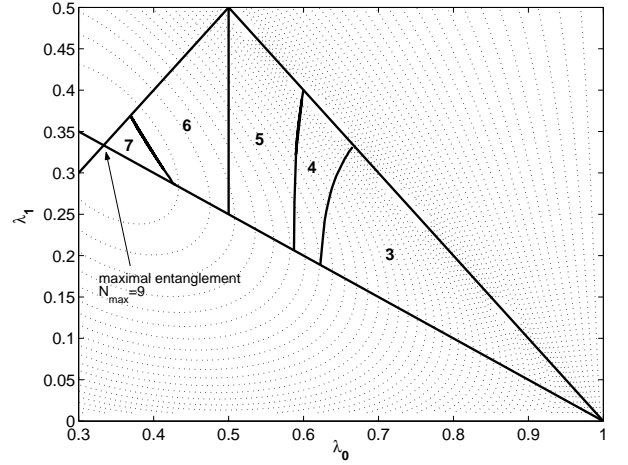}
\end{center}
  \caption[]{ Numerical mapping of $N_{max}(\psi)$, the maximal number of orthogonal unitaries
over the domain of pure states of 2 qutrits
$\ket{\psi} = \sqrt{\lambda_0}\ket{00} + \sqrt{\lambda_1}\ket{11} + \sqrt{1-\lambda_0-\lambda_1}\ket{22}$.
The horizontal axis is $\lambda_0$ and the vertical is $\lambda_1$.
The region of interest is defined by $\lambda_1 \le \lambda_0$,
$\lambda_0 + \lambda_1 \leq 1$ and $\lambda_1 \geq \frac{1-\lambda_0}{2}$.
This region is divided into 5 sub-regions characterized by different $N_{max}(\psi)$.
Contour lines of entanglement $S(\psi)$ are plotted in the background.
Note that no region with $8$
unitaries was found. The only case where $9$ unitaries exist is the maximally
entangled state ($\lambda_0=\lambda_1=\frac{1}{3}$).
}
\label{ent3d} \end{figure}

In the general case, we were unable to find
a parametrization of eq. (7) which leads to an analytic solution.
Numeric results are, however, obtainable.
In $d=3$ dimensions, using numeric root finding routines we have mapped the domain of pure states according to the
maximal number of orthogonal unitaries $N_{max}$ one can construct for a given initial pure state.
These results are presented in Figure \ref{ent3d}.
As we have already seen by example, we find that even when the initial entanglement
is less than one ebit, it is possible to construct more than
three orthogonal unitaries.

An intriguing observation is that we did not find partially
entangled pure states that enable the construction of a maximal set of orthogonal unitaries
of size $d^2-1$ (but we did find all steps $N_{max}\le d^2-2$).
Due to the increasing size of the numeric root finding problems, we have only been
able to verify this statement for $d=3,4$, and, of course, we have proved that this is
the case in two dimensions. If true, this is indeed a very peculiar property.

\begin{figure}
\begin{center}
\includegraphics[width=0.45\textwidth]{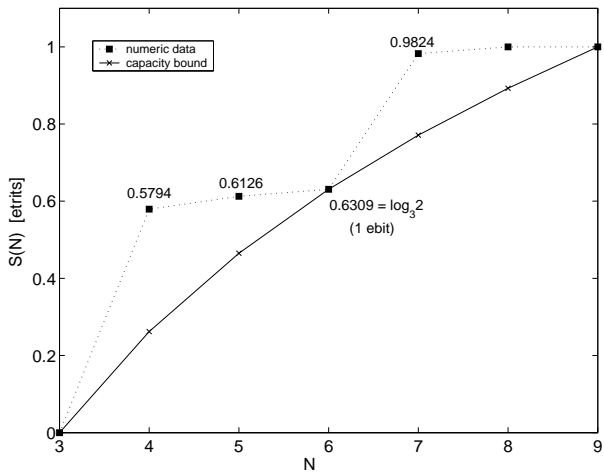}
\end{center}
  \caption[]{ Minimal entanglement (in etrits) required to construct $N$
orthogonal unitaries
as a function of $N$ in the 3-dimensional case.
Numeric results are marked by black squares connected with a dotted line.
The channel capacity bound is marked x connected by a solid line.
}
\label{minent} \end{figure}

It is also interesting to extract from the numerical results the minimal entanglement
necessary to construct $N$ orthogonal unitaries. One can compare this quantity with
the lower bound on the amount of entanglement derived from the channel capacity
 \cite{capacity}, which when measured in units of $d$its is given by
\beq
C \leq 1 + S(\psi)
\eeq
Therefore, the entanglement is
bound from below by $S(\psi) \ge log_dN -1$ e$d$its. Figure
\ref{minent} shows the comparison between the two quantities. It
is evident that only when $N$ is a multiple of $d$ do we achieve
this bound.

It is instructive to consider the states with minimal entanglement that enable the
construction of at least $N$ orthogonal unitaries. For $N<2d$ this entanglement is
less than one ebit. The states with this minimal entanglement have only two non vanishing
Schmidt coefficients, so they can be characterized by the value of $\lambda_0$ alone.
Table I shows $\lambda_0$ for different values of $N$ and $d$. Although this data
has been generated numerically, these values seem to be simple fractions
of the two quantities.
These results suggest that the state with minimal entanglement admitting at least $d+1$
orthogonal unitaries in $d$ dimensions is
$\sqrt{\frac{d-1}{d}}\ket{00}+\sqrt{\frac{1}{d}}\ket{11}$
(the states used (15)),
and that the state with minimal entanglement admitting at least $d+n$ ($n=2 \dots d$) orthogonal unitaries is
$\sqrt{\frac{d}{d+n}}\ket{00}+\sqrt{\frac{n}{d+n}}\ket{11}$.

\begin{table} \label{enttab}
\begin{tabular}{|c|c c c c c c c c c|}
\hline
& 2 & 3 & 4 & 5 & 6 & 7 &$\dots$ & $\dots$ & $d$ \\
\hline
3 & $- $ &  &  &  & & & & & \\
4 & $2/4$ & $2/3$ &  &  & & & & & \\
5 & & $3/5$ & $3/4$ & & & & & &\\
6 & & $3/6$ & $4/6$ &  $4/5$ & & & & &\\
7 & & & $4/7$ & $5/7$ & $5/6$ & & & &\\
8 & & & $4/8$& $5/8$ & $6/8$ &$6/7$ & & &\\
9 & & & & $5/9$ & $6/9$ &$7/9$ & & &\\
10 & & & &  $5/10$ & $6/10$ &$7/10$ & & &\\
11 & & & & & $6/11$ &$7/11$ & & &\\
12 & & & & & $6/12$ & $7/12$ & & &\\
13 & & & & & & $7/13$ & & &\\
14 & & & & & & $7/14$ & & &\\
$\vdots$ & & & & & & & $\ddots$& $\ddots$ & \\
$d+1$ & & & & &  & & & $\ddots$ & $\frac{d-1}{d}$ \\
$d+2$ &   & & & & & &  & & $\frac{d}{d+2}$ \\
$\vdots$ &   & & & & & &  & & $\vdots$ \\
$2d-1$  &  & & & & & &  & & $\frac{d}{2d-1}$ \\
$2d$  &  & & & & & &  & & $\frac{d}{2d}$ \\
\hline
\end{tabular}
\caption{Values of $\lambda_0$ for states with minimal entanglement
such that there exists a
construction of $N$ (row index) orthogonal unitary transformations in $d$ (column index)
dimensions. Numeric data and conjectured behavior are shown. Note that
the entanglement is smaller than one ebit.}
\end{table}

\section{Conclusions}
We have shown that \emph{deterministic} dense coding can be achieved using partially entangled
states.
While for qubits ($d=2$), partial entanglement does not help to improve the
classical communication capacity, in higher dimensions it is possible to have
deterministic dense coding even with less than one ebit (see Fig. 2).
When less than one ebit is shared by the parties,
the maximal number of alphabet letters that can be
communicated by a single particle is $2d-1$.
More generally, we show that the
alphabet size grows in ``steps"  and can obtain
the values $ d, d+1, \dots, d^2-2$ (Fig. 1).
We also find that
states with less entanglement can have greater communication
capacity than other more entangled states.

Table I summarizes the structure of states with minimal entanglement
smaller than one ebit, admitting $d<N<2d$ unitaries in $d$ dimensions.
The resulting structure strongly indicates
that geometric constructions, similar to the $N=d+1$ case (Section VI),
can be obtained as well.

A connection between superdense coding and other tasks such as teleportation and
distinguishability of operators has been noted in the past. In \cite{werner},
a one-to-one correspondence between dense coding schemes and
quantum teleportation schemes (for maximal entanglement) was established, and we
have already pointed out the similarity between the phase operators presented in section
\ref{geometric} and the teleportation protocol with partially entangled states discovered
independently in
\cite{G04}. It would be interesting to understand the correspondence between dense coding and
teleportation schemes when partial entanglement is used.

The problem of distinguishing unitary operators
and its relation to superdense coding in the maximal case
was presented in \cite{CPR00}. The conditions for distinguishing
a pair of unitary operations have been specified in \cite{acin}.
It is interesting that our constructions provide
non-trivial sets of unitary operators which can be perfectly distinguished
by a {\em single} measurement of a specific partially entangled state.

Finally, it would be interesting to examine whether the construction
of a the set of unitaries that satisfy the generalized orthogonality
condition (7), sheds light on the recent proposal for
probabilistic interpretation of evolutions \cite{unitary}.

\begin{acknowledgements}
SM and BR acknowledge the support of ISF grant 62/01-1.
JO is supported by EU grant
PROSECCO (IST-2001-39227), and a grant from the Cambridge-MIT Institute as
well as ISF grant 129/00-1.  Part of this research
was conducted during the Banasque session on Quantum Information
and Communication, 2003, and we thank the town and the organizers
for their hospitality.
\end{acknowledgements}

\end{document}